\documentstyle[prl,aps,multicol]{revtex}
\tighten
\input psfig
\begin{document}
\draft
\title{
Mesoscopic Fluctuations of Elastic Cotunneling
}
\author{I.L. Aleiner$^{*}$ and L.I. Glazman}
\address{Theoretical Physics Institute, University of
Minnesota, Minneapolis MN 55455}
\maketitle

\begin{abstract}
We study  mesoscopic fluctuations of the conductance through a quantum
dot at the wings of the  Coulomb blokade peaks.  At low temperatures,
the main mechanism of conduction is the elastic cotunneling. The
conductance  strongly fluctuates with an  applied magnetic field.  The
magnetic correlation field is shown to be controlled by the charging
energy, and the correlation function has a universal form. The
distribution function for the conductance  obtained analytically shows
a non-trivial crossover between the orthogonal and unitary ensembles.
\end{abstract}

\begin{multicols}{2}
Statistical theory of electron eigenfunctions in  complex sytems was
developed in the early fifties\cite{Porter} for the description of the
absorption spectra of large atoms. Soon it was realized that the same
random matrix (RMT) approach can be applied to  the excitation spectrum
of small metallic grains\cite{GorkovEliashberg}. An important
distinction should be made between the excitations that preserve the
number of electrons in the grain, and the excitations that change this
number. This distinction is most easily seen in the case where the
introduction of an additional electron causes charging of the grain
and, thus, is associated with the charging energy $E_c$. In this case,
the addition spectrum shows the gap of the width $E_c$, whereas the
spectrum of electron-hole excitations is controlled by typically much
smaller energy scale $\Delta$ (here $\Delta$ is the mean level spacing
of the grain).

The addition spectrum manifests itself in transport experiments on 
electron tunneling between two leads through a quantum dot. The
electrostatic potential of the dot, and therefore the energy of the
electron addition $E$, can be controlled by a capacitively coupled gate
electrode\cite{Review}.  At temperatures $T$ much smaller than
$E_c$, the presence of the gap results in  supression of the
conductance at almost all gate voltages. Only if the gate voltage is
tuned to one of the discrete points of charge degeneracy, i.e. $E =0$,
is this suppression lifted. This phenomenon is known as the Coulomb
blockade. Question arises whether and when the results of RMT (which
does not take into account any charging gap) can be applied to the
statistics of the conductance in the Coulomb blockade regime. Being
interesting on its own, this question is also relevant for the recent
experimental studies of the conductance fluctuations\cite{Marcus}. 
The answer to this question depends on how close the system is to the
charge degeneracy point. Right at the degeneracy point, and at low
temperatures $T
\ll\Delta$, the transport occurs by means of  resonant tunneling
through a single state.  Thus, the charging energy $E_c$ is irrelevant,
and RMT provides an adequate description of the distribution function
of the peak heights\cite{Stone,Alhassid}. Supersymmetry
calculation\cite{EfetovAdv} allows for a description of the
crossover\cite{Super}, with the magnetic field $B$, between  the
orthogonal  ($B=0$) and unitary ($B \to\infty $) ensembles.

Situation changes drastically if the system is tuned away from the
charge degeneracy point, $E > \Delta \simeq T$. In this case the
transport is due to the virtual transitions of an electron via 
excited states of the dot (so-called elastic
cotunneling\cite{AverinNazarov}); many levels  with energies exceeding
$E$ contribute to the tunneling. The superposition of a large number of
tunneling amplitudes changes the  properties of the conductance
fluctuations, which is studied for the first time in this Letter.

We will show, that the correlation function $C(\Delta B)$ for the
conductance in the co-tunneling regime is universal (i.e. all the
dependences for different $E$ can be collapsed to a single curve upon
 rescaling of the magnetic field);  magnetic correlation field
$B_c$ is inversely proportional to
$\sqrt{E}$, and thus the charging energy controls the fluctuations of
the elastic cotunneling. Furthermore, the functional form of $C(\Delta
B)$  is entirely different from  the known results\cite{Alhassid}.
Finally, we will find  the distribution function of the conductance
for all values of the magnetic field.

\begin{mathletters}
The quantum dot attached to two leads is described by the Hamiltonian:
\begin{equation}
\hat{H} = \hat{H}_L + \hat{H}_R + \hat{H}_D + \hat{H}_T,
\label{fullHamiltonian}
\end{equation}
where the Hamiltonians of the left (L) and right (R) leads are given by
\begin{equation}
\hat{H}_L = \sum_k \xi_k^la^{\dagger}_ka_k, \quad  \hat{H}_R = \sum_k
\xi_k^rb^{\dagger}_kb_k,
\label{leadsHamiltonian}
\end{equation}
 and
$\xi_k^{l,r}$ is the one electron energy measured from the Fermi level.
Hamiltonian of the dot $\hat{H}_D$ has the form\cite{Review}
\begin{equation}
\hat{H}_D = \sum_k \xi_k^Dc^{\dagger}_kc_k +
E_c\left(\hat{n}-{\cal N}\right)^2,\quad\hat{n}=\sum_kc^{\dagger}_kc_k,
\label{dotHamiltonian}
\end{equation}
where $\xi_k^D$ describes the one-electron spectrum of the dot, and the
second term in $\hat{H}_D$ corresponds to the charging energy, and
$E_c=e^2/2C$.  Here $C$ is the capacitance of the dot, and ${\cal N}$
is the conventional dimensionless parameter related to the gate
voltage $V_g$ by
${\cal N}=V_g/eC_g$, with $C_g$ being the gate capacitance.  The 
tunneling
Hamiltonian couples the leads with the dot, and it has the form
\begin{equation}
\hat{H}_T = \sum_{k,p} t^{l}_{kp} a^{\dagger}_k c_p +\sum_{k,p} t^{r}_{kp}
b^{\dagger}_k c_p  +h.c.
\label{tunnelHamiltonian}
\end{equation}
Operators $a$, $b$ and $c$ in
Eqs.~(\ref{leadsHamiltonian})-(\ref{tunnelHamiltonian}) are the
corresponding fermionic operators.
\end{mathletters}

If tunneling is weak, the charge of the dot $\hat{n}$ is quantized.
Obviously, degeneracy of the charging energy in
Eq.~(\ref{dotHamiltonian}) corresponds to half-integer values ${\cal
N}_m=m+\frac{1}{2}$ of the dimensionless gate voltage ${\cal N}$.

If $V_g$ is tuned away from  a degeneracy point, it takes a finite
energy $E$ to add one electron (or hole) to the dot,
\begin{equation}
E=E_c|{\cal N}-{\cal N}_m|, \quad |{\cal N}-{\cal N}_m|<1/2.
\label{energy}
\end{equation}
Positive (negative) values of ${\cal N}-{\cal N}_m$ correspond to the
electron (hole)-like lowest charged excitations.

We are considering the strong Coulomb blockade away from the
resonance. Thus, we employ  perturbation theory
in  the tunneling Hamiltonian (\ref{tunnelHamiltonian}). The lowest
nonvanishing contribution to the conductance  $G$ is
\begin{equation}
G=\frac{2\pi e^2}{\hbar}\sum_{k,p}\left|A_{kp}\right|^2
\delta(\xi_k^l)\delta(\xi_p^r).
\label{conductance}
\end{equation}
Amplitude $A_{kp}$ corresponds to the process in which an electron
(hole) tunnels from  state $k$ in the left lead into a virtual state
in the dot, and then it tunnels out to  state $p$ of the second lead.
This amplitude is given by
\begin{equation}
A_{kp}=\sum_q t^l_{kq}\left(t^r_{pq}\right)^*\frac{1}{|\xi_q|+E}
\theta \left[\xi_q ({\cal N}-{\cal N}_m)\right].
\label{amplitude}
\end{equation}
The denominator in Eq.~(\ref{amplitude}) corresponds to the energy of
virtual state $q$ involved in the cotunneling process and the
step-function $\theta(x)$ selects the dominating (electron or hole)
channel.

In the most realistic case\cite{Marcus,Chang} of point contacts,
Eqs.~(\ref{conductance}) and (\ref{amplitude}) may be further
simplified. The tunneling matrix elements $|t^{l,r}_{kq}|^2$ do not
depend on the indices $k,q$, and can be related to the conductances of
the point contacts, $G_{l,r}=(2\pi
e^2/\hbar)\nu_d\nu_{l,r}|t^{l,r}|^2$; here $\nu_{l,r,d}$ are the
emsemble-averaged densities of states in the leads ($l$, $r$) and dot
($d$) respectively. Using these definitions, substituting
Eq.~(\ref{amplitude}) into Eq.~(\ref{conductance}), and performing the
summation over $k$ and $p$ in Eq.~(\ref{conductance}), we find:
\begin{equation}
G=\frac{\hbar}{2\pi e^2}G_lG_r\left|F({\bf R}_l, {\bf R}_r)\right|^2.
\label{cond}
\end{equation}
The dimensionless function $F({\bf R}_l, {\bf R}_r)$ contains all the
information about elastic cotunneling through the dot between the point
contacts located at ${\bf R}_l$ and ${\bf R}_r$,
\begin{equation}
F({\bf R}_l, {\bf R}_r)=\frac{1}{\nu_d}\sum_q
\frac{\psi_q^*({\bf R}_l)\psi_q({\bf R}_r)}{|\xi_q|+E}
\theta \left[\xi_q ({\cal N}-{\cal N}_m)\right],
\label{F}
\end{equation}
where $\psi_q$ is the one electron wave function in the closed dot. It
is useful to rewrite $F$ in terms of the retarded and advanced
one-electron Green functions ${\cal G}^{R,A}$ of the dot,
\begin{equation}
F({\bf R}_l,{\bf R}_r)=\frac{1}{\nu_d}\int \frac{d\omega}{2\pi i}
\frac{{\cal G}^{A}_\omega-{\cal G}^{R}_\omega}{|\omega|+E}
\theta \left[\omega ({\cal N}-{\cal N}_m)\right],
\label{F1}
\end{equation}
where
\[
{\cal G}^{R,A}_\omega={\cal G}^{R,A}_\omega({\bf R}_l,{\bf R}_r) =
\sum_q
\frac{\psi_q^*({\bf R}_r)\psi_q({\bf R}_l)}{\omega - \xi_q \pm i0}.
\]
(We put $\hbar=1$ in all the intermediate calculations).

Equations (\ref{cond}) and (\ref{F1}) express the elastic cotunneling
conductance in terms of the exact electron wavefunctions of the dot.
These functions vary strongly when the magnetic field is applied to the
dot or the shape of the dot is changed. Thus, the conductance is a
random quantity and one should consider different moments of the the
conductance distribution function.   We will employ the ensemble
averaging, which is equivalent to the averaging over applied magnetic
field or over  the peak index $m$. According to Eqs.~(\ref{cond}) and
(\ref{F1}), averaged moments of the conductance are expressed in the
terms of the averaged product of the Green functions. It is well
known\cite{AronovSharvin}, that if the dot in the metallic regime (the
transport mean free path or the size of the dot is much larger than
the Fermi wavelength), and the relevant energies are much larger than
$\Delta$, these products can be related to the generalized classical
correlators -- diffuson ${\cal D}$ and cooperon ${\cal C}$:
\begin{mathletters}
\begin{eqnarray}
\langle{\cal G}^{R}_{\omega_1,B_1}({\bf r},{\bf s})
{\cal G}^{A}_{\omega_2,B_2}({\bf s},{\bf r})\rangle
= 2\pi\nu_d{\cal D}_{\omega_1-\omega_2}^{B_1,B_2}\left({\bf r},{\bf
s}\right),
\label{diffuson} \\
\langle{\cal G}^{R}_{\omega_1,B_1}({\bf r},{\bf s})
{\cal G}^{A}_{\omega_2,B_2}({\bf r},{\bf s})\rangle
= 2\pi\nu_d{\cal C}_{\omega_1-\omega_2}^{B_1,B_2}\left({\bf
r},{\bf s}\right),
\label{cooperon}
\end{eqnarray}
where $\langle\dots\rangle$ stand for the ensemble averaging perfomed
under the fixed magnetic fields $B_1,B_2$.  The averages of the type
$\langle G^RG^R\rangle$ and $\langle G^AG^A\rangle$ are much smaller
and can be neglected.  If the sample is dirty, so that the motion of
electrons in the dot is diffusive, the diffuson and cooperon
(\ref{diffusoncooperon}) satisfy the equations
\label{diffusoncooperon}
\end{mathletters}
\begin{mathletters}\label{dcequation}
\begin{eqnarray}
\left[-i\omega + \!D \left(\!-i\nabla_{\bf r}+
\frac{e}{c}{\bf A}^-({\bf r})
\right)^2\right]{\cal D}_{\omega}^{B_1,B_2}\!=
\! \delta\left({\bf r}-{\bf
s}\right),
\label{diffusonequation}\\
\left[-i\omega + \!D \left(\!-i\nabla_{\bf r}+
\frac{e}{c}{\bf A}^+({\bf r})
\right)^2\right]{\cal C}_{\omega}^{B_1,B_2}\!=
\!\delta\left({\bf r}-{\bf
s}\right),
\label{cooperonequation}
\end{eqnarray}
where $D$ is the diffusion constant, and ${\bf A}^\pm$ is the vector
potential due to the magnetic field, $\nabla\times {\bf A}^\pm = {\bf
B}_{1}\pm{\bf B}_{2}$. For the dot in the ballistic regime, the
diffusion operator in the l.h.s. of Eqs.~(\ref{dcequation}) should be
replaced with the Liouvillean operator. Solution of
Eqs.~(\ref{dcequation}) with the condition of vanishing normal
component of the gauge invariant current at the boundary of the dot
will enable us to find all the relevant correlation functions of the
conductance and we are turning to this calculation now.
\end{mathletters}

The averaged cotunneling conductance is obtained immediately by the
averaging of Eq.~(\ref{cond}) with the help of Eqs.~(\ref{F1}) and
(\ref{diffuson}). The result is
\begin{equation}
\langle G\rangle =\frac{G_lG_r}{2\pi^2\nu_d e^2}\int_{-\infty}^\infty
\frac{d\omega}{\left|\omega\right|}{\cal D}^{0,0}_{\omega}({\bf R}_l,
{\bf R}_r) \ln\frac{E+|\omega|}{E}.
\label{averagedconductance}
\end{equation}
Calculation of the second moment of the conductance is performed by
taking into  account that, because the relevant energy scale is much
larger than
$\Delta$, the average of the product of four Green functions can be
decoupled into a product of the pairwise averaged Green functions,
which are in turn given by Eqs.~(\ref{diffusoncooperon}). This yields
\begin{eqnarray}
&\displaystyle{\langle G(B)G(B+\Delta B)\rangle
=\left(\frac{G_lG_r}{2\pi^2\nu_d
e^2}\right)^2\times}&\label{correlator}\\
&\displaystyle{
\left[
        \left|
                \int_{-\infty}^\infty
                 \frac{d\omega}{\left|\omega\right|}
                  {\cal D}^{0,0}_{\omega}({\bf R}_l, {\bf R}_r)
                  \ln\frac{E+|\omega|}{E}
          \right|^2+
\right.}&
\nonumber\\
&\displaystyle{
\left.
          \left|
                 \int_{-\infty}^\infty
                  \frac{d\omega}{\left|\omega\right|}
                   {\cal D}^{B,B+\Delta B}_{\omega}({\bf R}_l, 
{\bf R}_r)
                 \ln\frac{E+|\omega|}{E}
            \right|^2 +
\right.}&\nonumber\\
&\displaystyle{
\left.
          \left|
                 \int_{-\infty}^\infty
                  \frac{d\omega}{\left|\omega\right|}
                   {\cal C}^{B,B+\Delta B}_{\omega}({\bf R}_l, 
{\bf R}_r)
                 \ln\frac{E+|\omega|}{E}
            \right|^2
\right].}&
\nonumber
\end{eqnarray}
The term in the second line of Eq.~(\ref{correlator}) corresponds to
the square of the average conductance, and the last two terms describe
the conductance fluctuations.  The term in the third line of
Eq.~(\ref{correlator})  depends only on $\Delta B$, [cf.
Eq.~(\ref{diffusonequation})] and it is present both for the
orthogonal  ($B=0$) and for the unitary ($B \to \infty$) ensembles. To
the contrary, the last term in Eq.~(\ref{correlator}) dies out for the
unitary ensemble.

It is easily seen from Eq.~(\ref{correlator}) that the fluctuations are
always of the order of the conductance itself.  This may be understood
from the following qualitative consideration. There are $N\sim
E/\Delta\gg 1$ contributions corresponding to different eigenstates  in
the cotunneling amplitude (\ref{amplitude}). Assume that the phases of
these contributions are completely random. Conductance is proportional
to the modulus squared of the sum of these contributions, and thus
there are $N^2$ terms in the conductance. Among those, $N$ terms do not
fluctuate, and the rest $N^2-N$ are random. These random terms,
however, do contribute to the fluctuation
$\langle \delta G^2\rangle$, and the number of non-vanishing terms in
it is $N^2-N$. Therefore, the average conductance is proportional to
$N$, and its r.m.s. fluctuation is $\sim\sqrt{N^2-N}\simeq N$. Thus,
conductance in the cotunneling regime is not a self-averaging quantity
despite a naive expectation that a large number of virtual states
participating in the cotunneling may decrease the fluctuations.

Equations (\ref{averagedconductance}) and (\ref{correlator}) are quite
general, i.e. they are valid for an arbitrary relation between the
 energy of charged excitation $E$ and Thouless energy $E_T\simeq\hbar
D/L^2$ (here $L$ is the linear size of the dot). In the most
interesting regime, $E < E_T$, the correlation function of conductance
fluctuations
$C(\Delta B)$ acquires  a universal form, as will be shown below.

Because $E < E_T$, the characteristic frequency $\omega$ in
Eqs.~(\ref{dcequation}) is much smaller than the lowest non-zero
eigenvalue of the diffusion operator (which is of the order of
$D/L^2$). Therefore, only the  zero frequency mode can be retained in
the solutions of Eqs.~(\ref{dcequation}). This mode corresponds to the
 probability density homogeneously distributed over the dot, and the
solution has the form:
\begin{eqnarray}
&\displaystyle{{\cal D}_\omega^{B_1,
B_2}=\frac{S^{-1}}{-i\omega +\Omega_-},
\quad
{\cal C}_\omega^{B_1, B_2}=\frac{S^{-1}}{-i\omega + \Omega_+};}&
\nonumber\\
&\displaystyle{\Omega_\pm=
E_T\frac{S^2\left(B_1\pm B_2\right)^2}{\Phi_0^2},}&
\label{zeromode}
\end{eqnarray}
where $S$ is the area of the dot, $\Phi_0 = 2\pi\hbar c/e$ is the flux
quantum, and the Thouless energy is given by $E_T= \alpha\hbar D/S$,
with   shape-dependent coefficient $\alpha$ of the order of unity.
Equation~(\ref{zeromode})  holds also for ballistic cavities; the only
difference is that the expression for the Thouless energy changes to 
$E_T\simeq\hbar/\tau_{fl}$, with $\tau_{fl}$ being the time of  flight
of an electron across the dot. Thouless energy can be independently
measured by studying the correlation function of mesoscopic
fluctuations for the same dot but with the contacts adjusted to the
ballistic regime.

Substitution of Eq.~(\ref{zeromode}) into
Eq.~(\ref{averagedconductance}) immediately yields  the
known\cite{AverinNazarov} result for the averaged conductance
\begin{equation}
\langle G\rangle =\frac{\hbar G_lG_r}{2\pi e^2}
\frac{\Delta}{E}.
\label{zeromodeconductance}
\end{equation}
However, the fluctuations $\delta G(B) = G(B) - \langle G\rangle$
are large. We find from Eq.~(\ref{correlator}) with the help of Eqs.
(\ref{zeromode}):
\begin{equation}
\frac{\langle\delta G(B) \delta G(B+\Delta B)\rangle}
{\langle G\rangle^2}\!
=\!
\Lambda\!\!\left(\frac{\Delta B}{B_c}\right)+
\Lambda\!\!\left(\frac{2B+\!\Delta B}{B_c}\right),
\label{zeromodecorrelator}
\end{equation}
where the scaling function $\Lambda(x)$ is given by
\begin{equation}
\Lambda (x)\! =\! \frac{1}{\pi^2\!x^4}
\!\!\left[\ln x^2\!\ln (1+x^4) +\pi
\arctan x^2\! +
\frac{1}{2} {\rm Li}_2(-x^4)
\right]^2\!\!\!\!,
\label{lambda}
\end{equation}
with ${\rm Li}_2(x)$ being the second polylogarithm
function\cite{Ryzhik}.
The asymptoic behavior of function $\Lambda$ is $\Lambda(x) =
1+(2x^2\ln x^2)/\pi$, for
$x
\ll 1$ and
$\Lambda(x) =(\pi x^2)^{-2} \ln^4x^2$ for $x\gg 1$.

The correlation magnetic field $B_c$ in Eq.~(\ref{zeromodecorrelator}) is
controlled by the charging energy
\begin{equation}
B_c = \frac{\Phi_0}{S}\sqrt{\frac{E}{E_T}}.
\label{Bc}
\end{equation}
It is worth noticing from Eq.~(\ref{zeromodeconductance}) and
Eq.~(\ref{Bc}) that the correlation magnetic field $B_c$ drops with
approaching a charge degeneracy point (in agreement with the recent
experiment\cite{Marcus}),  whereas the quantity $\langle G \rangle
B_c^2$ remains invariant. This invariance can be easily checked
experimentally.

Let us present also the expression for the experimantally measurable
correlation function of the conductance fluctuations $C(\Delta B)
= \langle\delta G(B)\delta G(B+\Delta B)\rangle/\langle\delta
G(B)^2\rangle$.
For both orthogonal and unitary ensembles we obtain from
Eq.~(\ref{zeromodecorrelator})
\begin{equation}
C(\Delta B) = \Lambda\left({\Delta B}/{B_c}\right),
\label{c}
\end{equation}
function $\Lambda(x)$ is defined by
Eq.~(\ref{lambda}). We emphasize  that the functional form 
of $C(\Delta
B)$ is different from the  results for the peak heights
fluctuations\cite{Alhassid}, see Fig.~\ref{fig1}.

{\narrowtext
\begin{figure}[h]
\vspace{-0.52cm}
\hspace*{-0.3cm}\psfig{figure=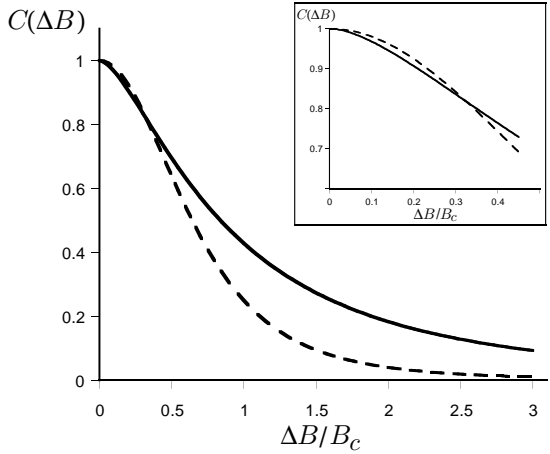,height=4.5in}
\vspace{-4.53cm}
\caption{The correlation function $C(\Delta B)$ for the conductance
fluctuations in the elastic cotunneling regime (solid line) and for
the peak height fluctuations $C=[1+(\Delta B/B_c)^2]^{-2}$ (dashed
line). For the elastic cotunneling, 
$C(\Delta B)$ is non-analytic at $\Delta B \to 0$, see the inset, and
Eq.~(\protect\ref{c}).}
\vspace{-0.22cm}
\label{fig1}
\end{figure}
}

As we saw, fluctuations of the conductance are of the order of the
averaged conductance. Thus the distribution function is non-gaussian,
i.e. it is not characterised by its second moment only. Fortunately,
for the elastic cotunneling the calculation of all the moments is
possible which enables us to find the distribution function $P(g)$
(here we introduced random variable $g=G(B)/\langle G\rangle$).
Function
$P(g)$ is defined as
\begin{equation}
\!\!P(g)\!\equiv \!\left< \delta \left( g -\! {{G} \over {\langle
G\rangle}}\right)\!\right>\! =\!
\int\!\frac{dq}{2\pi}e^{iqg}\!\sum_{n=0}^\infty
\frac{\left(-iq\right)^n\langle G^n\rangle}{n!{\langle G\rangle^n}}.
\label{Pdefinition}
\end{equation}
Calculation of the average $\langle G^n(B)\rangle$ entering into
Eq.~(\ref{Pdefinition}) is performed in a  fashion similar to
the derivation of Eq.~(\ref{correlator}). In the universal regime $E <
E_T$, we find
\begin{equation}
\frac{\langle G(B)^n\rangle}{\langle G\rangle^n}=
\sum_{0\leq j\leq n/2}
\frac{(n!)^2}{(j!)^2(n-2j)!}
\left(\frac{\lambda}{4}\right)^j,
\label{moments}
\end{equation}
where we introduced the short-hand notation $\lambda \equiv  \Lambda
(B/B_c)$. Substitution of Eq.~(\ref{moments}) into 
Eq.~(\ref{Pdefinition}) yields
\begin{equation} P(g) =
\frac{\theta(g)}{\sqrt{1-\lambda}}\exp\left(-\frac{g}{1-\lambda}\right)
{\rm I}_0\left(\frac{g\sqrt{\lambda}}{1-\lambda}\right),
\label{Presult}
\end{equation} 
where ${\rm I}_0(x)$ is the zeroth order modified Bessel
function of the first kind. In the limiting cases $\lambda = 1$ and
$\lambda =0$, the distribution function $P(g)$ coincides with the
Porter-Thomas distribution\cite{PorterThomas,Porter} for the
orthogonal and unitary ensembles respectively.

So far we considered the elastic cotunneling only. It dominates over
the inelastic processes\cite{AverinNazarov} at $T < \sqrt{E\Delta}$,
which is the typical regime for the modern experiments with
semiconductor quantum dots\cite{Marcus,Chang}. At higher temperatures,
the main conduction mechanism switches to the inelastic cotunneling.
Nevertheless, the fluctuations are still determined by the elastic
mechanism for
$\left(E\Delta\right)^{1/2}\lesssim T
\lesssim \left(E^2\Delta\right)^{1/3}$. At even higher temperatures,
the inelastic contribution dominates also in the fluctuations.
 Their relative magnitude, however, is small,
$<\delta G_{in}^2>/<G_{in}>^2 \simeq \Delta/T$. The correlation
magnetic field is controlled by the temperature rather than by the
charging energy and therefore is independent on the gate voltage.

In conclusion, we studied the statistics of  mesoscopic fluctuations of
the elastic cotunneling. We showed that the correlation magnetic field
is controlled by the charging energy and the correlation function of
the conductance is universal.

We are grateful to C.M. Marcus for illuminating discussions  and
reading the manuscript. This work was supported by NSF Grant
DMR-9423244.

\end{multicols}
\end{document}